\documentclass[twocolumn,epjc3,nospthms]{svjour3}

\usepackage{graphics}
\usepackage[usenames,dvipsnames]{color}
\usepackage{graphicx} 
\usepackage{dcolumn}
\usepackage{bm}
\usepackage{epsfig}
\usepackage{pdflscape}
\usepackage{subfigure}
\usepackage{multirow}
\usepackage{microtype}
\usepackage{booktabs}
\usepackage[colorlinks=true,citecolor=blue,filecolor=blue,linkcolor=blue,urlcolor=blue,pdftex]{hyperref}
\usepackage{rotating}
\usepackage{longtable}
\usepackage{lineno}
\usepackage{siunitx}
\usepackage{academicons}
\usepackage{comment}

\journalname{Eur. Phys. J. C}
\newcommand{\orcidicon}{\includegraphics[width=8pt]{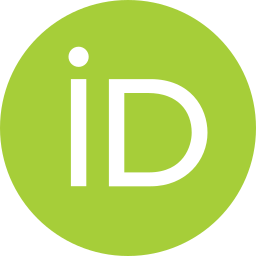}} 

\newcommand{\orcidlink}[1]{\href{https://orcid.org/#1}{\orcidicon}}
\begin{document}

\title{Improved and automated krypton assay for low-background xenon detectors with Auto-RGMS}

\author{
  Matteo~Guida\orcidlink{0000-0001-5126-0337}\thanksref{addr1,a}
  \and
  Ying-Ting~Lin\orcidlink{0000-0003-3631-1655}\thanksref{addr1}
  \and
  Hardy~Simgen\orcidlink{0000-0003-3074-0395}\thanksref{addr1}
}

\thankstext{a}{matteo.guida@mpi-hd.mpg.de (corresponding author)}

\institute{
  Max-Planck-Institut f\"ur Kernphysik, Saupfercheckweg 1, 69117 Heidelberg,
  Germany\label{addr1}
}

\date{Received: date / Accepted: date}

\maketitle

\begin{abstract}

Ultra-sensitive quantification of trace radioactive krypton-85 ($\textsuperscript{85}$Kr) is essential for low-background experiments, particularly for next-generation searches of galactic dark matter and neutrino physics using xenon-based time projection chambers (TPCs). While the rare gas mass spectrometer (RGMS) represents the current state-of-the-art for krypton detection in the field, we are developing a fully automated system (Auto-RGMS) to overcome the limitations of its manual operation. Auto-RGMS incorporates a robust control system for rapid measurements and minimized systematic uncertainties. A primary goal is to reach detection limits in the low parts-per-quadrillion (ppq) range for natural krypton by improving the chromatography stage to enhance the separation of krypton from xenon. Investigations into various adsorbent materials identified two candidates. HayeSep Q offers a 12-fold improvement in chromatographic resolution for xenon/krypton separation compared to the previously used adsorbent. Alternatively, HayeSep D provides a more limited improvement in resolution while allowing a higher measurement frequency because of its moderate retention-induced contamination after each measurement. By automating krypton assays and achieving ppq sensitivity, Auto-RGMS will be an indispensable tool for next-generation detectors, maximizing their scientific potential.
\end{abstract}

\section{Introduction}

Low-background noble gas detectors are widely used in particle physics for rare event searches, especially for direct dark matter searches. Liquid-xenon-based dual-phase time projection chambers (TPCs), such as XENONnT \cite{xenonnt}, LZ \cite{LZ:2022lsv}, and PandaX-4T \cite{pandax4t}, have currently achieved world-leading sensitivity in the searches for galactic weakly interacting massive particle (WIMP) dark matter. Meanwhile, liquid-argon-based experiments like DarkSide-50 \cite{darkside50} and DEAP-3600 \cite{deap-3600} offer complementary approaches. The next generation of experiments will be capable of exploring WIMP dark matter parameter space down to the neutrino fog \cite{PhysRevLett.127.251802}, while also driving progress in neutrino and Beyond Standard Model physics programs, creating opportunities for fundamental discoveries \cite{xlzd,xlzd_designBook}.

A common challenge in low-background experiments is radioactive contamination. Since noble gas targets are sourced from air, they inevitably contain trace impurities of other atmospheric components as a result of imperfect separation. Some of these impurities undergo radioactive decay, contributing to the detector's background noise. Even when considering commercially available high-purity xenon, which typically contains krypton at a concentration of 1-10 ppb, \textsuperscript{85}Kr remains present, representing a significant background in the dark matter search region of interest (ROI). To meet the stringent purity requirements for dark matter detection, a krypton reduction factor of at least $10^5$ must be achieved. 

The \textsuperscript{85}Kr isotope is a product of nuclear fission, introduced into the atmosphere through nuclear fuel reprocessing and past nuclear weapons testing. This results in an atmospheric activity concentration of $1.39(8)$ Bq/m\textsuperscript{3} in Europe today \cite{schlosser2020krypton85}, which translates to a \textsuperscript{85}Kr/\textsuperscript{nat}Kr ratio of $2.33(14) \cdot 10^{-11} \, \text{mol/mol}$. \textsuperscript{85}Kr beta-decays into \textsuperscript{85}Rb with an end-point energy of 687~keV.  While 99.56\% of the decay goes to the ground state, in 0.44\% of cases, the decay is followed by a delayed $\gamma$ emission due to de-excitation of \textsuperscript{85}Rb (J\textsuperscript{$\pi$} = 9/2\textsuperscript{+}) with a relaxation time of 1.015 $\mu$s \cite{SIEVERS1991271}. As a noble gas, \textsuperscript{nat}Kr is uniformly distributed in xenon TPC volume, making spatial discrimination techniques like shielding or fiducialization ineffective. For this reason, \textsuperscript{85}Kr is classified as an intrinsic background. In the XENON dark matter program, krypton impurities are reduced via a cryogenic distillation column. 
 The distillation method leverages the higher vapor pressure of krypton relative to xenon at -98°C, such that a single pass through the column achieves a krypton reduction factor of $(6.4^{+1.9}_{-1.4}) \cdot 10^5$ \cite{XENON_kr_2016}. The column operates at a rate of approximately 43 kg of xenon per day and supports both online and offline modes. In the offline mode, the entire xenon supply is distilled before entering the detector, while in the online mode, krypton is continuously removed during data acquisition. To date, the minimum krypton concentration observed was at the end of XENON1T, measured to be below 26 parts-per-quadrillion (ppq) \cite{XENON_kr_2016}.

The next generation of xenon-based dual-phase TPCs, such as DARWIN/XLZD, are targeting a \textsuperscript{nat}Kr/Xe concentration of 100 ppq for the WIMP dark matter search 
\cite{xlzd_designBook}. An additional major scientific goal of these experiments is to measure the solar proton-proton (pp) chain neutrino flux with unprecedented precision by utilizing elastic neutrino-electron scattering, thus surpassing the current best measurement from Borexino \cite{BOREXINO:2018ohr}. Since the shape of the solar pp neutrino energy spectrum, as observed via electron scattering, is nearly identical to that of \textsuperscript{85}Kr beta decays, \textsuperscript{85}Kr is one of the most critical background sources. Any suppression and more precise quantification of the \textsuperscript{85}Kr background will enhance the sensitivity. To reach the required flux measurement sensitivity, the \textsuperscript{nat}Kr/Xe ratio must be quantified with a precision on the order of a few ppq \cite{DARWIN:2020bnc}.

\section{Auto-RGMS: the automated rare gas mass spectrometer}
\label{sec2}

\subsection{RGMS: the rare gas mass spectrometer}

Up to now, krypton levels in xenon samples extracted from XENON dark matter program detectors have been quantified at the Max-Planck-Institut für Kernphysik (MPIK) in Heidelberg using the Rare Gas Mass Spectrometer (RGMS) \cite{hardy_lindemann}. Because feeding samples with predominantly xenon and trace amounts of krypton directly into a standard mass spectrometer would disrupt its vacuum conditions and damage the detector, the RGMS system consists of two stages for performing the measurement. 

The first stage employs gas chromatography to isolate krypton from xenon. This separation is achieved by adsorbent powder material with highly porous surfaces, which allows gas particles of different sizes to travel through at different velocities. The adsorbent material is housed in a U-tube-shaped column with an inlet and an outlet. When the sample is injected, krypton adheres less to the adsorbent than xenon and exits the outlet before xenon, achieving the separation. As carrier gas for sample transportation, purified helium is used, which is prepared by passing the already high-purity grade-6 helium through a Carbosieve S-III (Supelco, catalog 10184) adsorbent column cooled with liquid nitrogen. The chemical inertness of helium ensures that it freely passes through the adsorbent inside the columns. To prevent contamination, the system operates at an ultra-high vacuum with a leak rate less than $10^{-10}$~mbar$\cdot$l/s.

The second stage performs the counting measurement of the \textsuperscript{84}Kr and \textsuperscript{86}Kr natural isotopes via a customized version of a VG 3600 mass spectrometer \cite{VG3600}.  To calibrate the mass spectrometer, whose signal magnitude is proportional to the gas amount, a measurement of a gas mixture with precisely known amounts \cite{WeiserThesis} is performed before each measurement. We refer to this gas mixture as calibration gas. Moreover, blank measurements, where the full chromatography process is carried out without a sample, are performed to estimate the effect of out-gassing and residual gas in the RGMS system. With the sample krypton component isolated, signal size calibrated, and blank background subtracted, the RGMS measurement provides sample \textsuperscript{nat}Kr/Xe concentration with high sensitivity. Then with the knowledge of the atmospheric isotope abundance, the \textsuperscript{85}Kr concentration can be inferred.

The RGMS system had achieved the current world-best detection limit of 8~ppq of \textsuperscript{nat}Kr \cite{hardy_lindemann} in a xenon sample. However, this limit was obtained under the optimal condition, and due to practical limitations, maintaining such a condition over a long period of time is difficult, as will be discussed next.  

\subsection{Limitations of the RGMS performance}

While RGMS currently offers the world-leading krypton detection limit, it faces several challenges that demand improvements. The system was developed for isotopic studies in geosciences and is not optimized for trace impurity control. Long connection pipes and dead-ends in the tubing complicate the removal of residual krypton and xenon. Minuscule changes in leak tightness in any part of the large system impact the blank background level, making the 8~ppq limit not easily reproducible. Moreover, because the RGMS system operates manually, ensuring identical procedures for each measurement is challenging, which introduces systematic uncertainties and results in significant human resource costs. Operational errors can even result in significant downtime, which upon recovery, is often accompanied by increases in blank background levels. Finally, after each successful measurement, the system requires a one-workday-long bake-out and flushing with purified helium phase to clean the chromatography adsorbent from residuals of the previous sample. 

To achieve the demanding requirements for next-generation xenon observatories, an increase in the sample measurement frequency is required. That would allow for prompt detection of any sudden increase in krypton level, potentially caused by air leaks during xenon detector operations, and prevent detector downtime. Timely measurement will also be beneficial for characterizing the time-dependent variations in krypton concentration within the xenon detector. Furthermore, the next-generation experiment demands a system with detection limits in the ppq range for the solar pp neutrino measurement \cite{DARWIN:2020bnc}. This sensitivity improvement can be achieved by processing significantly larger xenon samples, which are not accessible with the separation ability of the current RGMS chromatography column.
 
\subsection{Auto-RGMS: the evolution of RGMS}
To overcome the issues discussed in the previous section, a fully automated Rare Gas Mass Spectrometer (Auto-RGMS) is being developed, which involves a completely newly built chromatography stage. The Auto-RGMS features a compact design that minimizes surface area and operates as an ultrahigh vacuum system, requiring the turbo molecular pumps to run continuously in a 24/7 mode.  All the vacuum pipes are maintained at elevated temperatures of up to 80$~^\circ\mathrm{C}$ to effectively remove residual gas and reduce redeposition. A getter pump is placed at the end of the chromatography stage to remove chemically active contaminants, keeping predominantly noble gases for the measurement. All valves in the system have fully welded bellows so that no potentially leaky seals are used. All fittings are face-sealed with metal gaskets. In most cases, 16~mm ConFlat flanges are applied. The design also incorporates multiple bypasses to the chromatography columns, providing significant debugging flexibility. Collectively, these features facilitate a clean and maintainable blank background level.

As indicated by its name, Auto-RGMS is designed to fully automate the measurement process once a sample is connected to the system. The automation interface is implemented using the LabVIEW software. A programmable XML recipe commands each operational step. The software parses the recipe, checks for errors, stores it in a database, and allows access through a recipe management system. The software program then executes each step while continuously monitoring and recording the evolution of system conditions, including temperature, flow rate, and pressure. These monitoring data, along with the measurement data from the mass spectrometer, are stored in a PostgreSQL database. An uninterruptible power supply (UPS) is installed to prevent disruptions from power outages.

\subsection{Feasibility studies for Auto-RGMS}

Several crucial improvements had to be demonstrated before the design of Auto-RGMS could be established. They involve:
\begin{enumerate}
    \item The development of a homemade all-metal double-valve for calibration purposes.
    \item The realization of a chromatography column design with thermal stabilization at both elevated and cryogenic temperatures.
    \item Stability tests of flow rate and pressure under realistic conditions.
    \item The selection of better-qualified adsorbents for the gas chromatographic Kr/Xe separation.
\end{enumerate}
For these studies, prototypes of the all-metal double-valve, the cooling/heating system for the chromatography column, and a small-scale demonstrator for the stability tests were built at MPIK. The same demonstrator was also used for the adsorbent tests. The R\&D efforts related to items 1–3, along with their results, are briefly summarized in this section, while the adsorbent selection process is discussed in greater detail in section~\ref{sec3}.

\subsubsection{All-metal double-valve}

\begin{figure}[!h]
\centering
\includegraphics[width=0.48\textwidth]{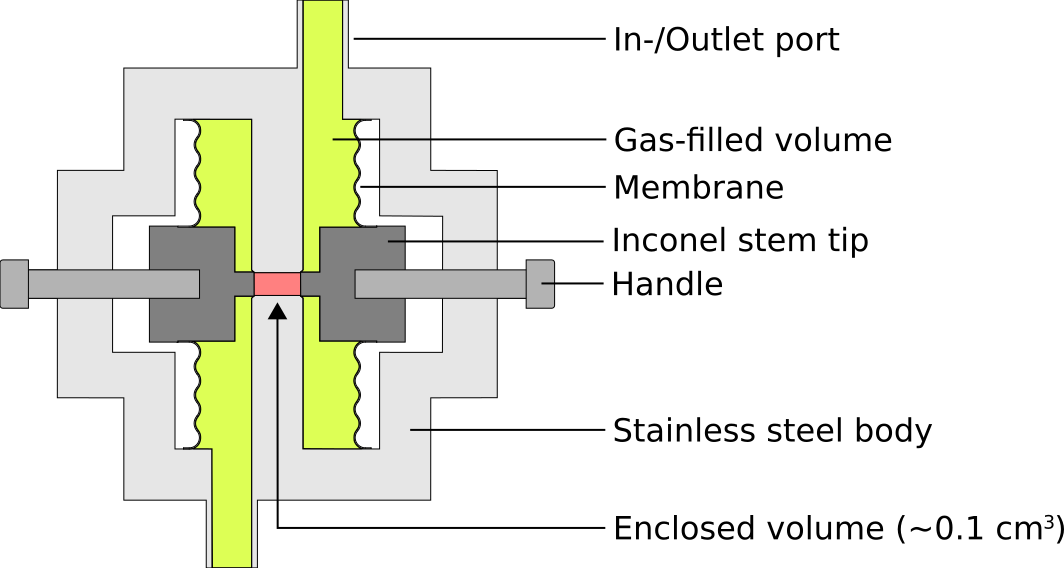}
    \caption{Sketch of the homemade double valve. Two stems with Inconel tips press on a stainless steel body, enclosing a volume of about 0.1 cm$^3$ (in red). The stems are connected to a membrane, providing necessary flexibility, and are operated by handles that apply consistent and uniform torque. For simplicity, only the right side of the double valve is labeled.\label{fig:double_valve}} 
\end{figure}

The response function of the mass spectrometer is calibrated by measuring a small but well-known amount of calibration gas of a few $10^{-11}$~cm$^3$ at standard temperature and pressure (STP). The calibration gas reservoir is a mixture of xenon, krypton, and argon with precisely known fractions; argon is used to calibrate the spectrometer's mass response. From the reservoir, small aliquots are taken, and therefore, a small volume in the range of 0.1~cm$^3$ enclosed by two all-metal valves is required. Commercial all-metal valves usually exhibit significantly larger volumes, so a homemade double-valve was built. A sketch of this custom-built double-valve is shown in figure~\ref{fig:double_valve}. Its volume was measured by the gas expansion method. Gas was enclosed at a precisely measured pressure in that volume. Then, the confined gas was released into a separate, larger, and well-known volume. The double-valve volume can be derived from the observed pressure drop. Using this technique, the prototype double-valve volume was measured as $0.0961(3)\, \text{cm}^3$. The custom-built double-valve is equipped with two motors for automated operation, ensuring a constant torque to prevent volume deformation. The electrical drives of the motors can be controlled remotely by LabVIEW software. Small volume shifts can be monitored and corrected by iterating the gas expansion method. Because the tightness of the passage cannot be guaranteed for the custom-built double-valve, two additional automated shut-off valves are fitted upstream and downstream.

\subsubsection{Chromatography column design}
\label{subsec_column_design}

The reproducibility of the chromatographic process is the most crucial aspect for reliable results with Auto-RGMS. It requires a well-controlled temperature of the adsorbent in the chromatography column in a temperature range from liquid nitrogen temperature (-196~$^\circ$C) to +200~$^\circ$C. This is achieved by immersing the U-tube-shaped stainless steel column containing the adsorbent powder in a liquid nitrogen bath with an automatic refill system. The stainless steel column is housed within a copper cylinder fitted with two 350 W heaters, ensuring uniform heat distribution. These heaters counteract liquid nitrogen cooling, ensuring precise temperature control up to the targeted separation temperature, e.g. from -90~$^\circ$C to 0~$^\circ$C as shown in table~\ref{tab:adsorbent_performance}. After separation, during the cleaning phase, the liquid nitrogen is evaporated, and the heaters raise the temperature to high levels (up to +200~$^\circ$C) to remove residual gases. However, due to gaseous nitrogen bubble formation on surfaces, stable operation is only possible if the liquid nitrogen is not in direct contact with the copper cylinder. Therefore, a 5~mm-thick polytetrafluoroethylene (PTFE) mantle was added to the copper cylinder as a thermal insulator. With this design, the column temperature could be stabilized with a precision better than 0.5~$^\circ$C \cite{hammann_2024_13946149}. Other Auto-RGMS adsorbent columns, such as the column for carrier gas purification, are either held at liquid nitrogen temperature or are heated at up to 200 $^\circ$C, making their temperature stabilization simpler.

\subsubsection{Auto-RGMS demonstrator}\label{subsec_demonstrator}

\begin{figure}[h]
    \centering
    \begin{minipage}[b]{0.45\textwidth}
        \centering
        \includegraphics[width=\textwidth]{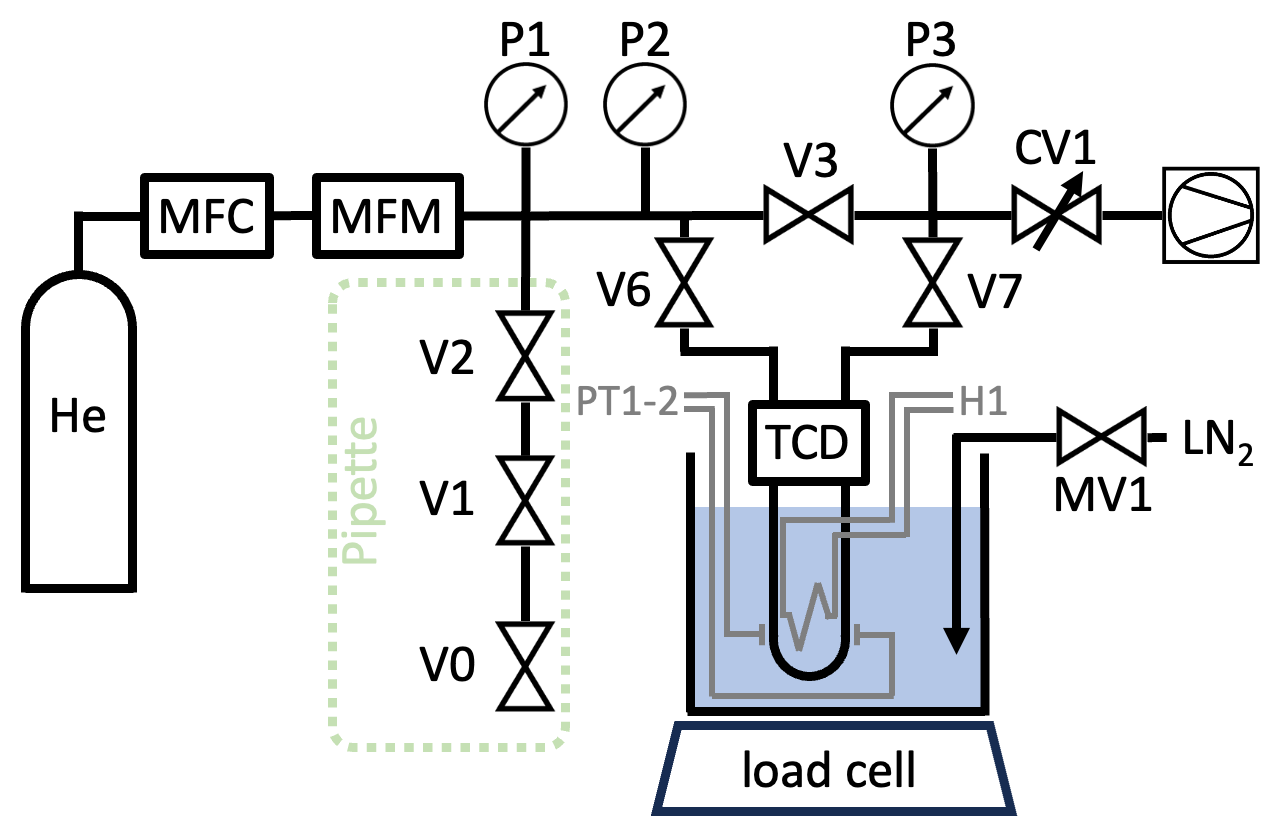}
    \end{minipage}
    \caption{Schematic of the Auto-RGMS demonstrator. \textit{He}: grade-6 helium bottle, \textit{MFC}: mass flow controller, \textit{MFM}: mass flow meter, \textit{P1}--\textit{P3}: pressure sensors, \textit{V0}--\textit{V7}: manual valves, \textit{CV1}: control valve, \textit{MV1}: magnetic valve, \textit{LN}$_2$: liquid nitrogen, \textit{TCD}: thermal conductivity detector, \textit{PT1}--\textit{PT2}: temperature sensors, \textit{H1}: heaters.}
    \label{fig:test_setup_combined}
\end{figure}

A prototype automated gas chromatography system was developed to validate the feasibility and automation concepts for the Auto-RGMS. Stability tests were conducted for critical parameters such as carrier gas flow rate, pressure, and column temperature. It consists of the key components sketched in figure~\ref{fig:test_setup_combined}.

\begin{table*}[htb]
    \renewcommand{\arraystretch}{1.2}
    \centering
    \label{tab:adsorbents}
    \begin{tabular}{@{} l S[table-format=4.0] S[table-format=3.0] S[table-format=2.1] S[table-format=1.1] @{}}
    \toprule
    {\small \textsf{\textbf{Adsorbent}}}       & {\small \textsf{\textbf{Surface Area [m\textsuperscript{2}/g]}}} & {\small \textsf{\textbf{Max Temp. [°C]}}} & {\small \textsf{\textbf{Avg. Pore Diameter [nm]}}} & {\small \textsf{\textbf{Density [g/mL]}}} \\
    \midrule
    Chromosorb 102           & 350                          & 230\textsuperscript{*}        & 8.5                             & 0.30                   \\
    HayeSep D                & 795                          & 230\textsuperscript{*}        & 30.0                            & 0.33                   \\
    HayeSep Q                & 582                          & 260\textsuperscript{*}        & 7.5                             & 0.35                   \\
    Shincarbon-ST            & 1500                         & 280                           & {0.5--1.5}                           & 0.51                   \\
    Silica MCM-48            & {1400--1600}                 & {$>$500}                      & 3.0                             & {0.5--1.0}             \\
    \bottomrule
\end{tabular}
\caption{Comparison of adsorbents from various suppliers: Chromosorb 102 (Chromatography Research Supplies, Inc.), Shincarbon-ST (Shinwa Chemical Industries LTD), HayeSep Q (Supelco, catalog 10300-u), HayeSep D (Supelco, catalog 10291), and Silica mesoporous MCM-48 (Sigma-Aldrich, catalog 805467-5G). The maximum temperatures listed with \textsuperscript{*} are not the manufacturers' specifications but rather the temperatures at which we observed the powders begin to exhibit a yellowish coloration; for more details, see section~\ref{memory}. Data also from \cite{bruno2020handbook} and \cite{analytics_shop}.}
\label{tab:adsorbents}
\end{table*}

\begin{itemize}

\item Temperature controlled adsorbent column: The column is loaded with the specific test adsorbent materials and follows the design explained in section~\ref{subsec_column_design}. For heating, two heaters (H1) are in place, while for cooling, an automated refill LN$_2$ dewar is controlled by a magnetic valve (MV1) and a load cell. The column temperature is read by two PT1000 sensors (PT1--2).

\item Sample injection system: A glass pipette with two precisely calibrated volumes (Vol$_1$ = 3.560(5) cm$^3$ and Vol$_2$ = 276.580(7) cm$^3$) 
and three manual valves (V0-V3) allows for controlled sample introduction.

\item Carrier gas supply: A high-purity grade-6 helium bottle provides a constant flow, regulated by a mass flow controller (MFC) and monitored by a mass flow meter (MFM).

\item Vacuum system: A scroll pump maintains the vacuum downstream. Two sensors P1 (full-range) and P2 monitor upstream pressure, while P3 monitors downstream pressure. A control valve (CV1) fine-tunes the outlet flow rate and pressure.

\item Detection system: A commercial thermistor thermal conductivity detector (TCD) by GOW-MAC is incorporated to detect the passing gases. The TCD features two thermistor-equipped tubes: one for inflow and one for outflow. While this detector is orders of magnitude less sensitive than the mass spectrometer of RGMS, it is significantly more flexible to operate and has sufficient sensitivity for the demonstrator phase. 

\item Data acquisition and control: A compact DAQ system interfaces with sensors and actuators, managed by a custom LabVIEW application.  
\end{itemize}

With the prototype automated setup, all three quantities temperature, pressure, and flow rate could be kept stable well below one percent for an operation time of more than 10 hours \cite{hammann_2024_13946149}. The prototype results under realistic operation parameters demonstrate the control feasibility of Auto-RGMS. This Auto-RGMS demonstrator is not only capable of performing stability tests, but it is also an excellent test bench for examining xenon-krypton separation for identifying the optimal adsorbent materials for Auto-RGMS. This will be discussed next.

\section{Chromatography adsorbents selection for Kr/Xe separation}\label{sec3}

As previously discussed, enhancing the RGMS's capability for \textsuperscript{nat}Kr detection in xenon requires improvements to the chromatography process. 
Increasing sample batch size directly scales with detectable krypton amounts, improving the detection limit. 
However, this approach depends on achieving robust separation between krypton and xenon. 
This is not feasible with the adsorbent currently used in RGMS due to its limited ability to separate the two gases. 
Specifically, the separation column must effectively resolve the chromatographic peaks of krypton and xenon to prevent xenon contamination in the mass spectrometer and ensure that the krypton tail is not prematurely truncated, which would result in krypton loss. 

A common approach to improve chromatographic resolution is to increase the length of the separation column, as resolution scales with the square root of its length. 
However, Auto-RGMS only tolerates relatively short columns due to its exceptionally stringent purity requirements.
We term the occurrence of high blank measurements after a measurement and cleaning procedure as the memory effect. Longer columns increase the impact of the memory effect, where trace gases, especially the dominant sample component xenon, become trapped within the adsorbent's pores, preventing complete removal. These trapped atoms significantly prolong the necessary cleaning process by baking, pumping, and flushing, potentially affecting subsequent analyses. 
Therefore, we searched for a better adsorbent material for our application accompanied by the optimization of operating temperature and helium flow rate to enhance separation performance and reduce the memory effect.

\subsection{Separation results}\label{subsec1}
Due to adsorption complexity, adsorbent choices are typically based on experiments instead of simulations. Moreover, data-sheet metrics such as surface area and pore size do not reliably indicate separation performance. Hence, application-specific experiments are essential to identify a suitable adsorbent.

We used the Auto-RGMS prototype presented in section~\ref{subsec_demonstrator} to test several porous solid materials as potential adsorbents for the Auto-RGMS system, aiming to identify a superior alternative to the Chromosorb 102 currently used in RGMS.  The key characteristics of these materials are summarized in table \ref{tab:adsorbents}. Chromosorb 102 is an organic compound with relatively large pores and we measured its performance as a reference for the other candidates. We tested two options from the HayeSep porous polymer family, HayeSep Q and HayeSep D, as they are commonly used as enhanced alternatives to Chromosorb 102 in gas chromatography applications and represent a more modern advancement in organic materials. While carbon molecular sieves are less common in chromatography, Shincarbon-ST is an exception, making it an interesting candidate. The inorganic silica MCM-48 was also tested for its potential stability and promising pore size. Organic materials are more prone to degradation and the outgassing of contaminants. In contrast, MCM-48's resistance to environmental factors, such as temperature and humidity fluctuations, allowed it to withstand higher baking temperatures.  

\begin{table*}[!ht]
\renewcommand{\arraystretch}{1.2}
\setlength{\tabcolsep}{8pt}
\centering
\begin{tabular}{@{} l c c | c c c c c c | c@{}}
\toprule
\textbf{Adsorbent} &
\begin{tabular}[c]{@{}c@{}}$\bm{T}_{\textbf{T}}$\\\textbf{[°C]}\end{tabular} &
\begin{tabular}[c]{@{}c@{}}$\bm{F}_{\textbf{T}}$\\\textbf{[sccm]}\end{tabular} &
\begin{tabular}[c]{@{}c@{}}$\bm{T}_{\textbf{S}}$\\\textbf{[°C]}\end{tabular} &
\begin{tabular}[c]{@{}c@{}}$\bm{F}_{\textbf{S}}$\\\textbf{[sccm]}\end{tabular} &
\begin{tabular}[c]{@{}c@{}}$\bm{t}^{\textbf{Kr}}_{\textbf{R}}$\\\textbf{[hh:mm:ss]}\end{tabular} &
$\bm{A}^{\textbf{Kr}}_{\textbf{S}}$ &
\begin{tabular}[c]{@{}c@{}}$\bm{t}^{\textbf{Xe}}_{\textbf{R}}$\\\textbf{[hh:mm:ss]}\end{tabular} &
\begin{tabular}[c]{@{}c@{}}$\bm{A}^{\textbf{Xe}}_{\textbf{S}}$\end{tabular} &
$\bm{R}$ \\
\midrule
Chromosorb 102    & -60 & 150 & -62(3) & 150.09(4) & 0:07:48(1) & 1.37(3) & 0:12:41(5) & 4.3(3) & 3.1(2) \\
                  & -75 & 150 & -76(1) & 149.96(5) & 0:09:41(1) & 1.42(3) & 0:19:40(10)& 9.0(6) & 3.7(3) \\
                  & -90 & 150 &  -90.0(3) & 149.79(8) & 0:08:49(2) & 1.6(5) & 0:43:20(80)& 7.7(2) & 4(1) \\
HayeSep D         & -60 & 150 & -60(1) & 150.53(3) & 0:10:17(1) & 1.25(3) & 0:28:00(10)& 3.7(2) & 4.5(3) \\
                  & -60 & 100 & -60.1(9) & 99.50(8) & 0:09:21(1) & 1.66(4) & 0:38:14(8) & 3.7(2)  & 5.7(2) \\
                  & -60 & 50  & -60.0(2) & 50.35(6)  & 0:16:06(2) & 1.32(6) & 1:10:50(30)& N/A & 6.6(6) \\
                  & -75 & 150 & -75.0(3) & 153.71(8) & 0:10:41(1) & 1.22(3) & 1:01:56(40)& 6.2(6) & 6.0(8) \\
                  & -90 & 150 & -89.9(2) & 149.90(9) & 0:15:00(3) & 1.31(5) & 3:22:00(160)& N/A & 8(1) \\
HayeSep Q         & -60 & 150 & -61(2) & 149.92(9) & 0:11:19(1) & 1.32(3) & 0:25:30(4)& 1.1(2) & 15(2) \\
                  & -75 & 150 & -75.3(4) & 150.10(3)  & 0:14:24(1) & 1.32(2) & 0:45:05.3(4)& 0.91(2) & 27.7(5) \\
                  & -90 & 150 & -90.0(2) & 151.91(5) & 0:12:53(2) & 1.41(4) & 1:43:19.3(3)& 1.27(3) & 51(1) \\
Shincarbon-ST     & 0   & 150 & 0(1) & 149.74(5) & 0:13:17(2) & 1.7(2) & 0:37:10(20)& 12(4) & 14(4) \\
Silica MCM-48     & -75 & 15  & -75.00(4) & 16(1) & 0:15:29(4) & 1.5(5) & 0:59:00(30) & 5.0(3) & 3.7(2) \\

\bottomrule
\end{tabular}

\caption{Performance metrics for various adsorbent materials evaluated for krypton/xenon separation using the Auto-RGMS demonstrator. Each measurement was conducted with approximately 2 grams of adsorbent material. The table lists the target temperature ($T_{\text{T}}$), target helium carrier gas flow rate ($F_{\text{T}}$), the measured flow rate ($F_{\text{S}}$), and the measured temperature ($T_{\text{S}}$) during the separation. Additionally, it presents the retention times for krypton ($t^{\text{Kr}}_{\text{R}}$) and xenon ($t^{\text{Xe}}_{\text{R}}$), the asymmetry factors for krypton ($A^{\text{Kr}}_{\text{S}}$) and xenon ($A^{\text{Xe}}_{\text{S}}$), and the resulting chromatographic resolution ($R$). Uncertainties in the operational parameters represent the sample standard deviation of their values between the krypton and xenon peaks.
}
\label{tab:adsorbent_performance}
\end{table*}

For each experiment, approximately 2 grams of adsorbent material is loaded into the stainless steel U-tube.
Small plugs of glass wool were inserted on either side of the powder to secure its position. The gas sample used has a much larger krypton concentration, Kr:Xe ratio of 1:10, than in a typical sample measured by the RGMS. This way, the sensitivity of the TCD is sufficient to detect both components of the gas mixture. The amount of sample gas employed in each experiment was between 2--3~cm\textsuperscript{3}~ STP.
To assess the performance of different adsorbents in krypton/xenon separation, systematic tests were carried out under identical experimental conditions. During these experiments, the helium carrier gas flow rate was maintained at 150 sccm.

\begin{figure}[!h]
\centering
\includegraphics[width=0.48\textwidth]{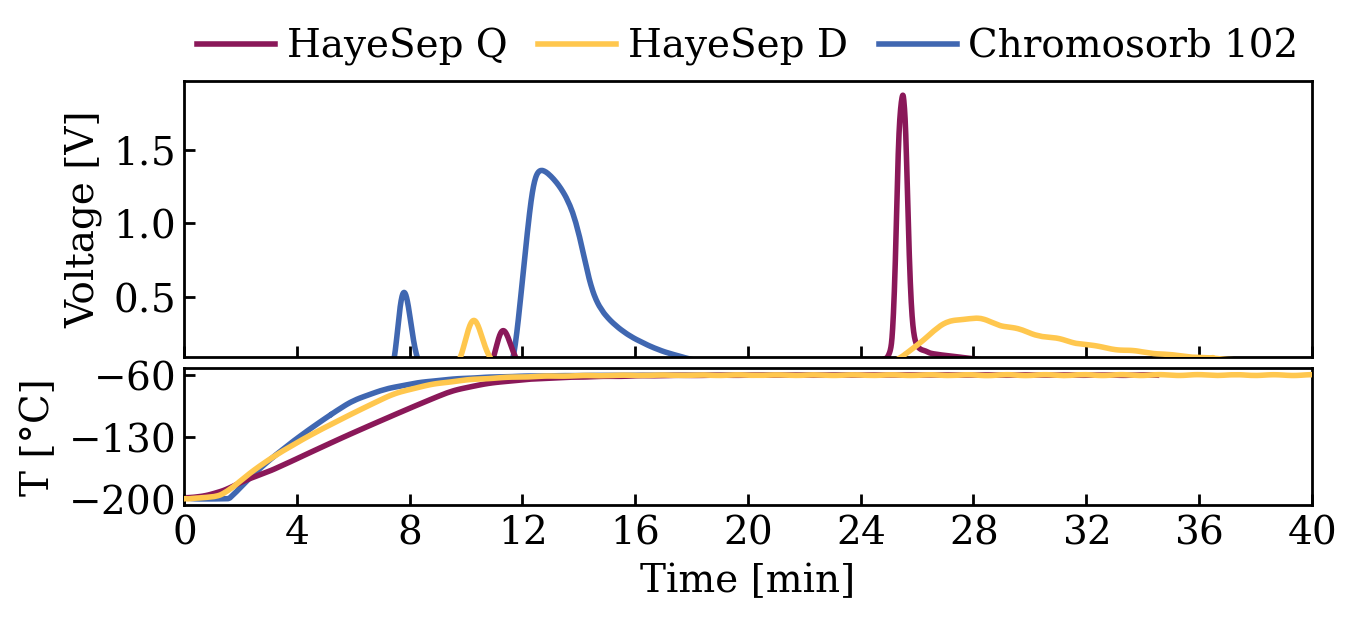}
    \caption{Chromatogram of krypton and xenon elution profiles for different adsorbents (HayeSep Q, HayeSep D, Chromosorb 102) at -60°C, acquired with the Auto-RGMS demonstrator.
The upper panel shows the thermal conductivity detector (TCD) signal corresponding to the thermistor voltage. 
The lower panel displays the column temperature during separation, starting from -196\textdegree C (liquid nitrogen) and ramping up to the target temperature of -60\textdegree C.\label{fig:chromatogram_runs_21_37_28}} 
\end{figure}

Figure~\ref{fig:chromatogram_runs_21_37_28} shows typical chromatograms obtained with HayeSep Q, HayeSep D, and Chromosorb 102. These chromatograms illustrate the elution profiles of krypton and xenon. The first peak tracks the krypton, the second one to xenon with the peak positions and shapes reflecting the separation characteristics of each adsorbent. The chromatograms also provide valuable information on peak symmetry and tailing, which greatly influence the performance of the separation. Significant tailing of the xenon peak manifests on Chromosorb 102 and HayeSep D. The underlying cause of this phenomenon may involve xenon's distinctive interaction characteristics with these adsorbents, which exhibit stronger and more heterogeneous binding sites compared to krypton \cite{Fornstedt1996PeakTA}, as well as column overloading, where xenon operates beyond its linear regime \cite{dettmer2014practical}.
In our application, xenon peak tailing was found to correlate with larger sample sizes, along with a decrease in retention time \cite{winkler2020,hammann_2024_13946149}. Meanwhile, krypton displayed stable peak symmetry and position. This is consistent with the expectation that symmetric distributions are not affected by concentration increases in terms of peak displacement \cite{grob2004modern}.
Although theoretically reducing the sample size would help reestablish the quasilinear regime, this approach is infeasible for our application, as we aim in Auto-RGMS to enhance the sensitivity by increasing the sample size.

The principal metric for evaluating separation performance is the chromatographic resolution (\( R \)) between the krypton and xenon peaks. The resolution is a quantitative measure of the separation between two chromatographic peaks, incorporating both the difference in retention times and the extent of peak broadening. A higher \( R \) value indicates improved separation efficiency. For asymmetric peak shapes, it is defined as \cite{dettmer2014practical}:

\[
R = \frac{t^{\mathrm{Xe}}_{\mathrm{R}} - t^{\mathrm{Kr}}_{\mathrm{R}}}{1.7\left(w^{\mathrm{Kr}}_{\mathrm{tail}} + w^{\mathrm{Xe}}_{\mathrm{front}}\right)} ,
\] 
\vspace{1mm} 
where $t^{\mathrm{Kr}}_{\mathrm{R}}$ and $t^{\mathrm{Xe}}_{\mathrm{R}}$ denote the retention times of krypton and xenon, respectively, and $w^{\mathrm{Kr}}_{\mathrm{tail}}$ and $w^{\mathrm{Xe}}_{\mathrm{front}}$ represent the widths at half-maximum of the tail and front of the corresponding peaks. The factor of 1.7 accounts for the use of half-maximum widths rather than full widths at half-maximum. To evaluate the symmetry of the peaks, the asymmetry factor ($A_s$) is used, calculated as:

\[
A_s = \frac{b}{a} ,
\] where $a$ and $b$ are the front and back half-widths of the peak at 10\% of its height.

\begin{figure}[!h]
\centering
\includegraphics[width=0.48\textwidth]{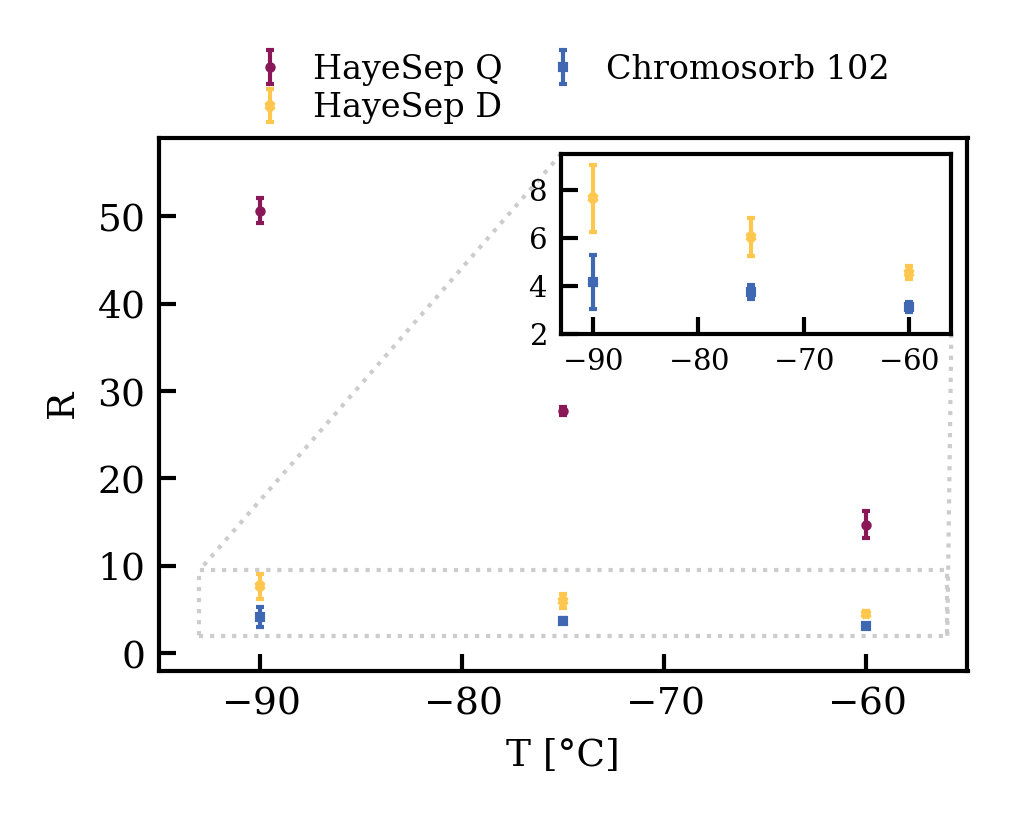}
\caption{Chromatographic resolution (R) of krypton and xenon separation across various temperatures for the adsorbents listed in table~\ref{tab:adsorbents}. Shincarbon-ST and Silica MCM-48 are
is excluded as it was evaluated under different experimental conditions. Its performance is discussed in the text.\label{fig:R_vs_temperature_with_inset}}  
\end{figure}

The resolution measurement results for the tested adsorbents are presented in table~\ref{tab:adsorbent_performance} and figure~\ref{fig:R_vs_temperature_with_inset}. The figure illustrates the temperature-dependent behavior and the variation among different adsorbents. HayeSep Q not only exhibits better performance at a particular temperature, but its performance enhancement when the temperature is reduced is also substantially superior to that of the other adsorbents. This is due to the fact that HayeSep Q is still in linear regime for our sample size resulting in greater symmetry and sharpness of the xenon peak, which is noticeably visible in figure~\ref{fig:chromatogram_runs_21_37_28}.


The performance of Shincarbon-ST, tested at a higher temperature with satisfactory results already observed, is presented in table~\ref{tab:adsorbent_performance}. However, memory effect measurements, which are discussed in section~\ref{memory}, excluded the material for Auto-RGMS. For this reason, no further measurements at lower temperatures were taken. The commercially available grain size of Silica MCM-48 rendered it unsuitable, as it could not maintain the standard helium flow rate of 150 sccm without causing a severe pressure increase before the column. This necessitated a drastic reduction to 15 sccm to operate under comparable pressure conditions to the other adsorbents (see table~\ref{tab:adsorbent_performance}). Furthermore, its separation performance, even at this lower flow rate, was poorer than the other tested powders.

\begin{figure}[h]
\centering
\includegraphics[width=0.48\textwidth]{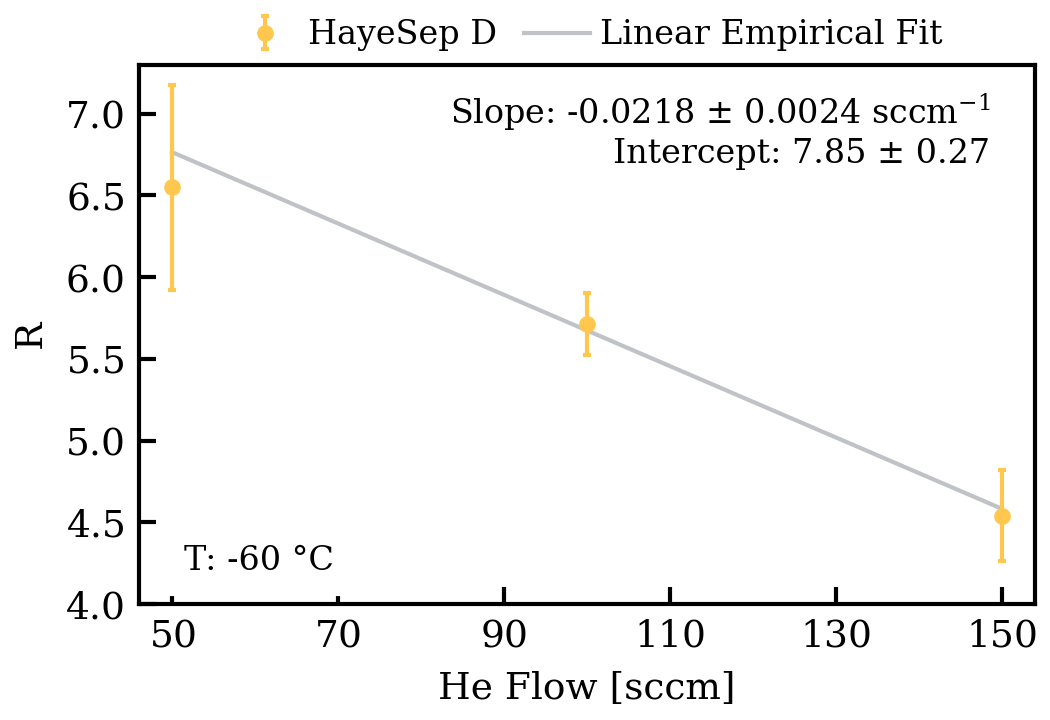}
\caption{Chromatographic resolution (R) of krypton and xenon separation for HayeSep D as a function of helium flow rate at -60 °C. \label{fig:r_vs_he_flows_hayesepd}}
\end{figure}

Helium flow rate significantly influences separation performance. Lower flow rates improve chromatographic resolution but lead to longer measurement time. Figure~\ref{fig:r_vs_he_flows_hayesepd} shows the linear decrease in resolution observed for HayeSep D at \(-60 \, ^\circ\mathrm{C}\) with increasing helium flow. In this specific case at a helium flow rate of 150 sccm, the maximum of the xenon peak occurs after approximately 28 minutes. Reducing the flow rate to 50 sccm shifts the peak to 1 hour and 11 minutes, which remains within an acceptable duration for our purposes. The helium flow rate for Auto-RGMS will be determined based on the final operational and performance requirements, while striving to keep it as low as possible to enhance chromatography resolution.

\subsection{Memory effect validation}
\label{memory}

In standard RGMS operations, residual gas is removed from the adsorbent by baking, pumping, and flushing with purified helium after each sample measurement. 

The chosen baking temperature was different for the different adsorbents. In particular, the organic HayeSep materials are delicate as they suffer from strong outgassing at elevated temperatures. With excess baking, they also tend to change their color from white to yellow. While it has been stated that the change of color does not affect the chromatography property \cite{bruno2020handbook}, we argue that it indicates changes in chemical composition, which is undesirable given the high purity requirement of our system. Accounting for all the factors, we baked HayeSep~D/Q to the temperature of 110$^{\circ}$C / 130$^{\circ}$C, respectively. ShinCarbon-ST is a pure carbon material, so we applied a higher temperature of 250$^{\circ}$C. We did not perform any memory effect tests for the MCM-48.

To best assess the impact of the memory effect, we directly performed the measurement using RGMS. We connect the same adsorbent column that was used for the separation tests directly to RGMS at an existing spare port at the end of the chromatography stage. Then we cleaned the adsorbent by baking and pumping only. Flushing was not applicable as the spare port did not provide access to purified helium. The results obtained are thus conservative because additional helium flushing will lead to substantially greater suppression of the memory effect.

We perform a blank measurement after the baking and pumping to ensure the column reaches the desired low background level. Next, we cool the column to liquid nitrogen temperature and expose the adsorbent to the RGMS calibration gas that contains a well-known amount of krypton, xenon, and argon \cite{WeiserThesis}. After one hour of collection, we isolate and heat the column for another hour to fully spread the calibration gas across the adsorbent. To simulate the typical cleaning process, we open the column to pump and bake for a day. Next, we close the column and accumulate the outgassing from the remaining traces of calibration gas for 40 minutes, the same collection time used for measuring XENON samples. Finally, the collected gas is analyzed with the RGMS. If the measurement result remains elevated, it indicates a non-negligible memory effect. In that case, we continue baking and pumping the column until the next measurement is conducted to study the extent of the memory effect.

A challenge of this procedure is to leak-check and pump down the adsorbent from air exposure to a blank background level sufficient for RGMS standards. Moreover, the memory effect measurements could only be performed between scheduled RGMS measurements for the XENONnT experiment, making the data points scarce. 

\begin{figure}[h]
\centering
\includegraphics[width=0.48\textwidth]{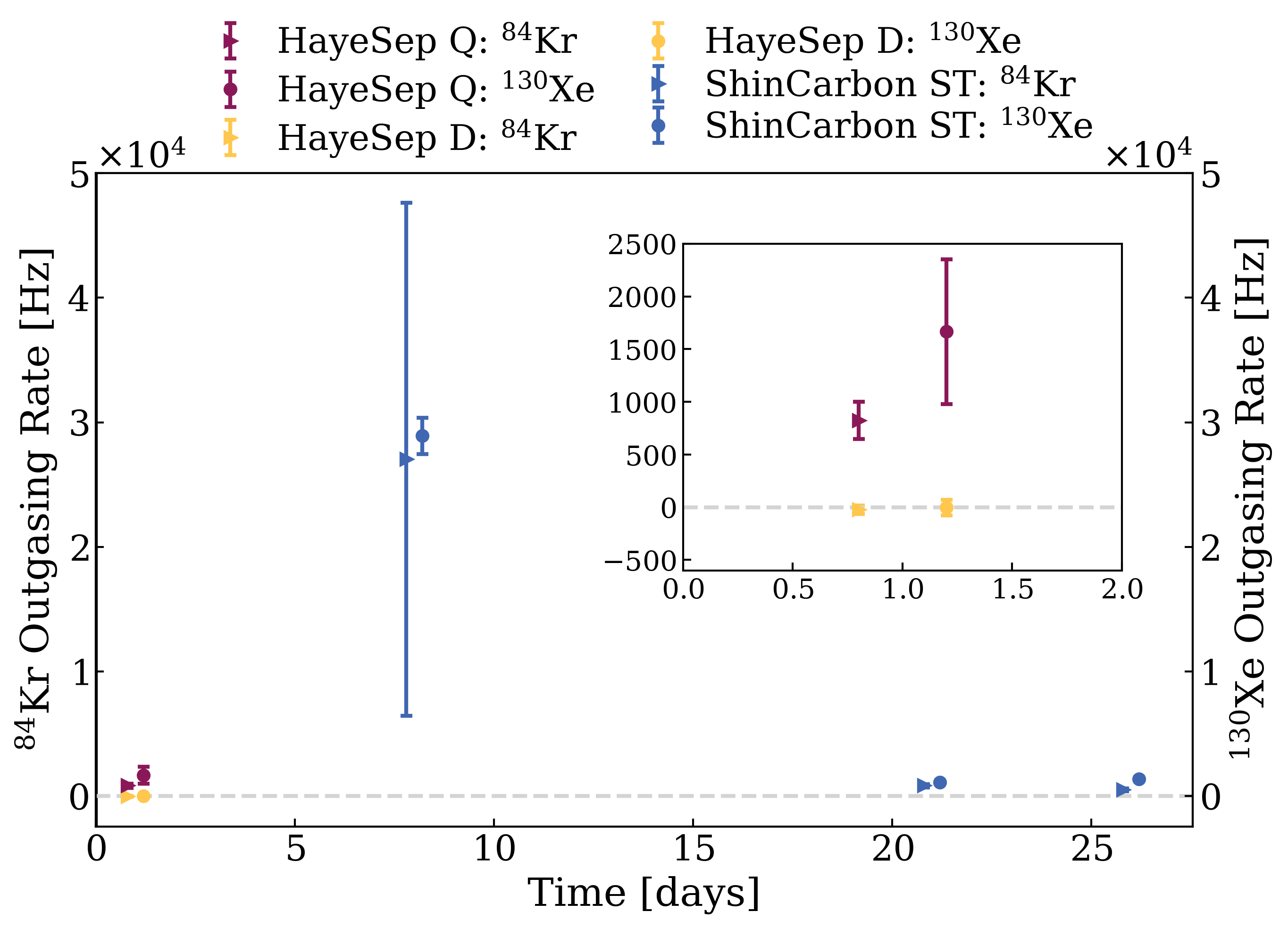}
\caption{Memory effect measurements. The purple data points show the results from HayeSep Q, yellow for HayeSep D, and blue for ShinCarbon ST. The triangles represent $^{84}$Kr results from the RGMS mass spectrometer measurement on the left y-axis, and the circles represent $^{130}$Xe results on the right y-axis. The x-axis shows the day of the heat and pumping process since the calibration gas injection. The offsets on the x-axis between the $^{84}$Kr and $^{130}$Xe results are for visibility.}
\label{fig:memory_effect}
\end{figure}

The blank-subtracted results of these measurements for HayeSep Q, HayeSep D, and ShinCarbon-ST as a function of pumping time are shown in figure~\ref{fig:memory_effect}. The measurement for ShinCarbon-ST after the first day was out of range for RGMS and was aborted to avoid damaging the detector of the mass spectrometer. The error of each data point is derived from a conservative estimate of the systematic to the detector response for each measurement and the systematic over the time period for each adsorbent. 

The measurement of $^{84}$Kr is representative of a typical RGMS measurement. If $^{84}$Kr cannot be cleaned after each sample measurement, the krypton level will be significantly biased for subsequent measurements. Considering that xenon is the most abundant gas in a sample, the measurement of $^{130}$Xe is through its doubly ionized channel that has a significantly lower signal rate to protect the detector. In other words, if xenon cannot be cleaned, it could potentially cause damage to the detector due to its heavy masses. Moreover, if substantial xenon gas enters the mass spectrometer, it will distort the krypton signal. This known detector effect is indicated by the large error bar of the ShinCarbon-ST data point in figure~\ref{fig:memory_effect}. The memory effect of both krypton and xenon will be problematic for sample measurements.

The results show that ShinCarbon-ST exhibits a substantial memory effect. The ShinCarbon-ST adsorbent is known to have a high fraction of micropores and does not achieve the desired blank background level even after a month of baking and pumping. This removes ShinCarbon-ST as a candidate for the Auto-RGMS. In contrast, the results of HayeSep~D show a negligible memory effect. As for HayeSep~Q, although the measurement was cut short due to a change in RGMS condition, the results show a noticeable memory effect, which undermined the otherwise great performance of HayeSep~Q.

As mentioned, the memory effect tests without helium flushing are conservative. In Auto-RGMS a continuous supply of purified helium is guaranteed, so helium flushing will be applied whenever an adsorbent column is prepared for the next measurement. We have performed a test to flush Hayesep~Q for one workday with a temporary supply of purified helium, and the result demonstrates that its memory effect can be sufficiently suppressed to a negligible background level. With the continuous flushing of Auto-RGMS, we can foresee reaching the blank background level in a shorter time. While the experimental conditions were not ideal, the test shows that HayeSep~Q is the most promising choice for Auto-RGMS.

\section{Conclusions}
We presented the development of Auto-RGMS, an automated rare gas mass spectrometer designed to precisely measure trace krypton in xenon-based detectors utilized in dark matter and neutrino research. Auto-RGMS significantly upgrades its manually operated predecessor by fully automating operational procedures and ensuring precise control over critical parameters such as temperature and carrier gas flow. Advancements include a custom all-metal double-valve for accurate calibration gas delivery and enhanced gas handling. Combined with a compact, optimized surface design, these features allow Auto-RGMS to reduce systematic uncertainties, stabilize blank backgrounds, and significantly reduce human labor. Improved sensitivity will be achieved by increasing sample volumes, which requires superior chromatographic resolution for effective krypton-xenon separation. 

Among the tested adsorbents, HayeSep~D provided a moderate enhancement in separation performance and, advantageously, showed no memory effects, thus facilitating quicker baseline recovery through simple pumping and baking. HayeSep~Q, on the other hand, was identified as the best-performing adsorbent, delivering a remarkable 12-fold enhancement in chromatographic resolution over the previous Chromosorb 102, allowing detection at the challenging low ppq level. The memory effect of HayeSep Q will be suppressed by using helium flushing in Auto-RGMS. Currently, the Auto-RGMS is being commissioned at MPIK, bringing it closer to use in future xenon observatories.

\section*{Acknowledgments}
We gratefully acknowledge the support of the technical facilities of the Max-Planck-Institut f\"ur Kernphysik. We are particularly indebted to Steffen Form, Klaus J\"anner, Michael Reissfelder and Jonas Westermann for their dedicated work and technical support during the construction and operation of the Auto-RGMS and demonstrator systems. We also thank Robert Hammann for valuable discussions. M.G. acknowledges the financial support of the International Max Planck Research School for Precision Tests of Fundamental Symmetries (IMPRS-PTFS). This work was supported by the Max Planck Society.
  
\bibliographystyle{utphys}
\bibliography{auto-rgms-paper.bib}

\end{document}